
-----------------------------------------------------------------------
The article is in plain TeX format.
The figures are attached following the manuscript.
They can be printed directly (postscript format).
-----------------------------------------------------------------------
\overfullrule=0pt
\def\jpa #1 #2 #3 {{\sl J. Phys.\ A} {\bf #1}, #2 (#3)}
\def\jcp #1 #2 #3 {{\sl J.\ Chem.\ Phys.} {\bf #1}, #2 (#3)}
\def\jdep #1 #2 #3 {{\sl J.\ de Physique I} {\bf #1}, #2 (#3)}
\def\pra #1 #2 #3 {{\sl Phys.\ Rev.\ A} {\bf #1}, #2 (#3)}
\def\prb #1 #2 #3 {{\sl Phys.\ Rev.\ B} {\bf #1}, #2 (#3)}
\def\pre #1 #2 #3 {{\sl Phys.\ Rev.\ E} {\bf #1}, #2 (#3)}
\def\prl #1 #2 #3 {{\sl Phys.\ Rev.\ Lett.} {\bf #1}, #2 (#3)}
\def\gtwid{\mathrel{\raise.3ex\hbox{$>$\kern-.75em\lower1ex\hbox{$\sim$}}}}
\def\ltwid{\mathrel{\raise.3ex\hbox{$<$\kern-.75em\lower1ex\hbox{$\sim$}}}}
\def\eg{{\it e.\ g.}}\def\ie{{\it i.\ e.}}

\def\pd#1#2{{\partial #1\over\partial #2}}      
\def\p2d#1#2{{\partial^2 #1\over\partial #2^2}} 
\def\ith{{$i^{\rm th}$}}

\newcount\refnum\refnum=0  
\def\refi{\smallskip\global\advance\refnum by 1\item{\the\refnum.}}
\newcount\rfignum\rfignum=0  
\def\rfigi{\medskip\global\advance\rfignum by 1\item{Figure \the\rfignum.}}
\def\pvt{{P(v,t)}}
\def\pinf{{P_{\infty}(v)}}
\def\pxvt{{P(x,v,t)}}
\def\pxvi{P_0(x,v)}
\def\pnvt{{P_m(v,t)}}
\def\pnt{{P_m(t)}}
\def\qnvt{{Q_m(v,t)}}
\def\pvi{{P_0(v)}}
\def\pv{{\psi(V)}}
\def\pn{{\phi(M)}}
\def\pnv{{\Phi(M,V)}}
\def\pvim{{P_0(v)}\sim v^{\mu}}
\def\svt{S(v,t)}
\def\rvt{R(v,t)}

\def\pg{P_+}
\def\ps{P_-}
\def\pvp{P_0(v')}

\def\ct{c(t)}
\def\cxt{c(X,t)}
\def\cinf{c_{\infty}}
\def\a{{\alpha}}
\def\b{{\beta}}
\def\m{\mu}
\def\cta{{c\sim t^{-\a}}}

\def\vtb{{v\sim t^{-\b}}}
\def\dx{\Delta x}
\def\dv{\Delta v}
\def\mo{(\m+1)}
\def\mt{(\m+2)}
\def\va{\mo/\mt}
\def\vb{{1/\mt}}
\def\apb{{\a+\b=1}}

\def\nav{\langle m(v)\rangle}
\def\lr#1#2#3{\left#1 #2 \right#3}
\def\l{\lambda}
\def\e{\epsilon}
\magnification 1200
\baselineskip=18 true bp
\centerline{\bf Kinetics of Clustering in Traffic Flows}
\bigskip
\centerline{\bf E.~Ben-Naim, P.~L.~Krapivsky and S.~Redner}
\smallskip
\centerline{Center for Polymer Studies and Department of Physics}
\centerline{Boston University, Boston, MA 02215}
\vskip 1in
\centerline{ABSTRACT}
{\smallskip\noindent We study a simple aggregation model that mimics the
clustering of traffic on a one-lane roadway.  In this model, each
``car'' moves ballistically at its initial velocity until it overtakes
the preceding car or cluster.  After this encounter, the incident car
assumes the velocity of the cluster which it has just joined.  The
properties of the initial distribution of velocities in the small
velocity limit control the long-time properties of the aggregation
process.  For an initial velocity distribution with a power-law tail at
small velocities, $\pvim$ as $v \to 0$, a simple scaling argument shows
that the average cluster size grows as $n \sim t^{\va}$ and that the
average velocity decays as $v \sim t^{-\vb}$ as $t\to \infty$.  We
derive an analytical solution for the survival probability of a single
car and an asymptotically exact expression for the joint mass-velocity
distribution function.  We also consider the properties of spatially
heterogeneous traffic and the kinetics of traffic clustering in the
presence of an input of cars.
}

{
\narrower\bigskip\noindent
P. A. C. S. Numbers: 68.70+w, 03.20.+i, 05.20.Dd, 05.40.+j.
}

\vfill\eject

\medskip\centerline{\bf I. Introduction}\smallskip

A variety of approaches have been applied to describe the collective
properties of traffic flows [1]. For example, to mimic congested traffic
flow in two dimensions, cellular automation models have been proposed
[2,3].  Asymmetric hopping processes have also been applied to model
traffic flow on a one-dimensional road [4-6].  When the number of cars
is large, traffic flows can be modelled phenomenologically in terms of a
one-dimensional compressible gas [7-9].  Such an approach predicts the
appearance of shock waves, where hydrodynamic quantities, such as the
average density and velocity, become discontinuous.  However, the
hydrodynamic approach does not naturally describe the behavior of
traffic flows in the low-density limit where there are large
heterogeneities in traffic density.  For this situation, a microscopic
model may provide a more appropriate description.

In this article, we introduce a ballistic aggregation process to model
the kinetics of clustering in one-dimensional traffic flows.  Our
approach is inspired, in part, by the recent interesting results that
have been obtained for a variety of reaction processes which involve
ballistic particles including: ballistic agglomeration, $A_i+A_j \to
A_{i+j}$, with momentum conserving collisions [10,11]; ballistic
annihilation, $A+A \to 0$, [12,13]; and several nucleation and ballistic
growth processes [14-16].  In our model, cars move ballistically in a
one direction, say to the right, according to an initial velocity
distribution.  Clusters form whenever a faster car overtakes a slower
car or cluster. The overtaking car then assumes the
velocity of the lead car in the cluster.  This model is an idealized
description for one-lane traffic flow.  While there are obvious
shortcomings in our model, it is exactly soluble and permits a thorough
understanding of the kinetics of the aggregation process.

This paper is organized as follows.  In section II, we present the model
and postulate the scaling behavior for the velocity and the
concentration of the clusters.  This approach makes use of the
statistical properties of the minimal random variable within a large
sample.  In section III, we investigate the distribution of cluster
velocities.  For this distribution, the cluster size is irrelevant and
this feature allows us to consider a simpler ``coalescence only'' model.
For this reduced problem, the velocity distribution is obtained exactly
in terms of the initial distribution of car velocities and then
evaluated for general continuous distributions.  Building on these
results, the general clustering process is solved in section IV and an
asymptotically exact expression for the joint cluster mass-velocity
distribution is obtained.  In section V, we present a formal solution
for the velocity distribution function for an inhomogeneous initial
distribution of particles. We examine  the temporal behavior
that arises for a simple step function initial spatial distribution.
  In section VI, we investigate another
generalization of the model to the situation with a spatially and
temporally homogeneous input of cars.  Depending on the functional form
of the input velocity distribution in the low-velocity limit, the input
can give rise to a steady state or to a system which continues to evolve
indefinitely.  We give our conclusions in section VII.  The details of
specific calculations are given in the Appendices.

\bigskip\centerline{\bf II. Scaling Analysis}\smallskip

We consider an idealized one-dimensional traffic flow in which the size
of each car is zero.  This is appropriate for describing clustered
traffic in the low-density limit, a situation which is often encountered
on rural secondary roads.  In the following, we will refer to such
sizeless cars as particles.  We consider the initial condition
when there are only isolated particles (``monomers'') in the system with
a random spatial distribution of density $c_0$. The initial velocity
distribution, $P(v,t=0)$, can generally be written in the scaling form
$$
P(v,t=0)={c_0\over v_0}P_0\lr ( {v\over v_0} ) \qquad v>0,
\eqno(1)
$$
with $\int_0^{\infty}P_0(z)dz=1$.  Here we have tacitly subtracted the
assumed finite value of the velocity of the slowest car from all
velocities.  In what follows, it is often convenient to introduce the
dimensionless density $c/c_0\to c$, velocity $v/v_0\to v$ and time
$c_0v_0t\to t$.  This yields a rescaled initial concentration which is
equal to unity.

In our model, particles move at their initial velocities and whenever a
particle overtakes a cluster aggregation occurs.  The aggregation rule
is simply that two colliding clusters form a new cluster with a velocity
equal to the smaller of the two incident cluster velocities and with a
mass equal to the sum of the two cluster masses (Figure 1).  If we
denote a cluster of mass $m$ and velocity $v$ by $A_{m,v}$, the process
is described by the reaction scheme
$$
A_{m_1,v_1} + A_{m_2,v_2}\to A_{m_1+m_2,{\rm min}\{v_1,v_2\}}.
\eqno(2)
$$

We present now a simple argument, based on the statistics of extremes,
to predict the asymptotic time dependence of the typical cluster mass
$m$ and typical cluster velocity $v$ at time $t$.  Since the typical
distance, $l$, between clusters grows with time as $l\sim vt$, the
typical number of particles in a cluster is proportional to this
distance, yielding $m \sim l \sim vt$. To find the typical velocity, one
has to relate the mass of a cluster to its velocity.  Such a relation
may be found exactly for an auxiliary ``one-sided'' problem in which
particles are placed with a fixed density $c_0$ to the left of a given
``target'' particle which moves at velocity $v$, and no particles are
placed to the right.  Eventually, this target particle
will form a cluster that includes all consecutive particles to its left
whose initial velocities are larger than $v$.  The probability that
there are exactly $k$ such particles is equal to $\ps\pg^k$, with
$\pg(v)$ $\bigl(\ps(v)\bigr)$ defined as the probability that a particle
has a velocity larger (smaller) than $v$,
\ie, \hbox{$\pg(v)=\int_v^\infty P_0(v')dv'$}.  Therefore, the average
number of particles in the cluster that ultimately forms is given by
$$
\nav=\sum_1^{\infty} k\ps\pg^k=\pg/\ps. \eqno(3)
$$

Let us now assume a power-law behavior of the initial velocity
distribution for small velocities,
$$
\pvi\simeq a v^{\m}, \qquad v\ll1
\eqno(4)
$$
with $\m>-1$ for normalizability.
Imposing this power-law form in Eq.~(3) yields
$$
\nav={\pg(v)\over\ps(v)}\propto {1\over v^{\mu+1}}, \eqno(5)
$$
for sufficiently low velocities.  For a particle which moves with the
typical velocity, it is reasonable to expect that this result for the
``one-sided'' problem gives the correct behavior for the full
``two-sided'' problem.  If we combine Eq.~(5) with our previous estimate
$m\sim vt$, we find the following asymptotic relations,
$$
\eqalign{m&\sim t^{\a}\cr v&\sim t^{-\beta}},\qquad
\eqalign{{\rm with}\cr{\rm with}}\qquad
\eqalign{\a&={\mu+1\over\mu+2}\cr\b&={1\over\mu+2}}.
\eqno(6)
$$

Since the mass is conserved in the aggregation process, the typical
cluster mass and the concentration of clusters $c$ are related by $c\sim
1/m\sim t^{-\a}$.  Notice that in the limit $\mu\to \infty$, the mass
grows linearly with time.  In contrast, when $\mu\to -1$, the mass is
roughly constant, since the velocity distribution becomes effectively
unimodal and collisions are exceedingly rare. This qualitative
dependence on the form of the initial velocity distribution is
reminiscent of the ballistic annihilation process [13], where
ballistically moving particles annihilate upon collision.  In both
processes, one finds that the fundamental exponents are related by
$\apb$ as a consequence of the relation $c\sim 1/vt$.  Moreover, for the
two processes the decay exponents have similar functional dependences on
the form of the initial velocity distribution.  However, the values of
the decay exponents are different: for example, for a uniform
distribution (corresponding to $\m=0$) one finds $\a=1/2$ for the
traffic model, while $\a\cong0.76$ is obtained in simulations of the
annihilation process.  Additionally, despite the qualitative similarity
between these two models for continuous velocity distributions,
different behaviors occur when the velocities are discrete.  For such
discrete distributions, the concentration typically decays algebraically in
time for ballistic annihilation [13], while the concentration decays
exponentially in time for the traffic model.

Since both the typical mass and the velocity scale as power laws in time,
the probability of finding a cluster of mass $m$ and velocity $v$,
$\pnvt$, is expected to evolve toward a scaling distribution.  Taking
into account mass conservation,
$\int_0^{\infty} dv\sum_m m \pnvt={\rm const.}$,
we postulate the scaling form
$$
\pnvt\simeq t^{\b-2\a}\pnv, \eqno(7)
$$
for the joint distribution, where the scaled mass, $M$, and scaled
velocity, $V$, are defined by
$$
M=m/t^{\a}\qquad{\rm and}\qquad V=vt^{\b}.\eqno(8)
$$
Note that while the mass $m$ is a discrete variable, the rescaled mass
$M$ is continuous.

Once the joint mass-velocity distribution
is found, the single-variable mass and velocity distributions can be
found by suitable integrations over the subsidiary variable.
Thus the velocity distribution, $\pvt=\sum_m\pnvt$, should have
the scaling form
$$
\pvt\simeq t^{\b-\a}\pv,\eqno(9a)
$$
with $\pv=\int_0^{\infty}dM\pnv$,  while the cluster-mass
distribution, $\pnt=\int_0^{\infty} \pnvt$, should have the scaling form
$$
\pnt\simeq t^{-2\a}\pn,\eqno(9b)
$$
with $\pn=\int_0^{\infty}dV\pnv$.

\bigskip\centerline{\bf III. The Car Survival Probability}\smallskip

As a preliminary step in obtaining a full solution for traffic
clustering, consider the velocity distribution $\pvt$.  For this
quantity, we can ignore the masses of each cluster
and focus only on the survival probability of a given car.  Thus  the
evolution of the velocity distribution is governed by the ``derived''
coalescence process
$$ A_{v_1}+A_{v_2}\to A_{{\rm
min}\{v_1,v_2\}}.\eqno(10)
$$
In the coalescence process, the density of particles with velocity $v$
is identical to $\pvt$, the velocity distribution of clusters in the
full traffic aggregation model defined by Eq.~(2).

Let $\svt$ be the survival probability of particles of velocity $v$ at
time $t$. Here ``survival'' means a car does not overtake any traffic,
but an overtaken car is defined to have survived.  Then the velocity
distribution function is given by
$$
\pvt=\pvi \svt. \eqno(11)
$$
The survival probability $\svt$ can be found by considering the possible
collisions of a particle with initial position and velocity $(x,v)$ with
slower particles whose initial positions are to the right of $x$.  A
collision between the initial particle with co-ordinates $(x,v)$ and a
slower $v'$-particle does not occur up to time $t$ if the interval
$[x,x+(v-v')t]$ does not include the slower particle.  For an initial
velocity distribution, $\pvi$, and a Poissonian initial spatial
distribution, the probability that there is no particle with velocity
between $v'$ and $v'+dv'$  in the interval $[x,x+(v-v')t]$ is
$$
\exp\lr [ {-dv'\pvp (v-v')t} ].
\eqno(12)
$$

For a particle to survive to time $t$, this exclusion probability should
be taken into account for every $v'<v$.  To verify this, let us assume
otherwise and derive a contradiction.  Thus consider a particle with
initial data $(x,v)$ that has maintained its original velocity to time
$t$.  In addition, assume that a slower $v'$-particle is initially
present in the above exclusion zone, \ie, $\dx(0)<\dv(0)t$. Here
$\dx(t)$ is the distance between the two particles and $\dv(t)$ the
relative velocity at time $t$. Since the velocity $v'$ can only decrease
over time due to collisions, one has $\dv(t)\ge\dv(0)$.  Consequently,
at time $t$, the separation between the two particles,
$\dx(t)=\dx(0)-\int_0^t\dv(t')dt'\le\dx(0)-\dv(0)t< 0$.  Thus the
$v$-particle does not survive, in contradiction with the original
assumption.

Hence, the survival probability is simply a product of the exponential
factors of Eq.~(12) for all $v'$, with $v'<v$.  Evaluating this product
gives the survival probability
$$
\svt=\exp\lr [ {-t\int_0^v dv'(v-v')\pvp} ] ,\eqno(13)
$$
and combining with Eq.~(13) yields the velocity distribution,
$$
\pvt=\pvi\exp\lr [ {-t\int_0^v dv'(v-v')\pvp} ] .\eqno(14)
$$
This is valid for an {\it arbitrary} initial velocity distribution
$\pvi$; the only source of stochasticity arises from the initial
conditions. For discrete initial distributions it is seen from Eq.~(14)
that the approach to the final concentration is exponential in time.
Thus, we focus only on the more interesting continuous initial velocity
distributions.

For the power-law initial velocity distribution $\pvi\simeq a v^{\m}$
for $v\ll1$, a direct calculation shows that the long-time velocity
distribution approaches a form that is independent of the details of the
large-velocity tail of the initial distribution,
$$
\pvt\simeq a v^{\m}\exp\lr [ {-b tv^{\m+2}} ] ,
\eqno(15)
$$
with $b=a/\mo\mt$.  This expression can be written in the scaling form
$(9a)$ with the scaling function
$$
\pv=a V^{\m}\exp\lr [ {-b V^{\m+2}} ] .
\eqno(16)
$$
{}From this solution we see that the velocity distribution maintains the
original power-law form for small velocities.  The exact solution also
validates the scaling assumption that the asymptotic decay as well as
the shape of the limiting distribution are determined solely by the
low-velocity tail of the initial distribution which, in turn, is
governed by the exponent $\m$.

{}From Eq.~(15), it is straightforward to compute the total concentration
and the average cluster velocity,
$$
\ct=\int_0^{\infty}dv\,\pvt, \quad
\langle v(t)\rangle= {1 \over \ct}\int_0^{\infty}dv\, v\, \pvt.
\eqno (17)
$$
This gives
$$
\ct\simeq \mo b\Gamma(\a)(b t)^{-\a}, \eqno(18)
$$
and
$$
\langle v(t)\rangle\simeq {1\over\Gamma(\a)}(b t)^{-\b}, \eqno(19)
$$
respectively. These expressions confirm the scaling laws suggested in Eq.~(6).

Interestingly, the exact solution of Eq.~(14)
satisfies the following Boltzmann-like integro-differential equation,
$$
\pd \pvt t=-\pvt\int_0^v dv'(v-v')P(v',0).
\eqno(20)
$$
This equation suggests that the loss of $v$-particles due to collisions
with slower $v'$-particles occurs at a rate proportional to the relative
velocity, $(v-v')$. Moreover, the pair correlation function factorizes
into a product of single-particle velocity distributions,
$P(v,v',t)=\pvt P(v',0)$ but with different time arguments for the two
factors.  In contrast, in the conventional Boltzmann equation, the
decomposition would involve the same argument for each velocity
distribution. Thus, the exact Eq.~(20) quantitatively indicates the
degree of approximation of the mean-field Boltzmann equation.

\bigskip\centerline{\bf IV. The Full Probability Distribution}\smallskip

We now solve for the joint mass-velocity distribution function for the
general traffic model.  To obtain $\pnvt$, the density of clusters of
mass $m$ and velocity $v$, it is useful to introduce the cumulative
distribution, $\qnvt$, the distribution of clusters of velocity $v$ and
mass greater than or equal to $m$.  Once the latter distribution is
known, $\pnvt$ can be obtained by
$$
\pnvt=\qnvt-Q_{m+1}(v,t).\eqno(21)
$$
Notice that the density of clusters of mass greater or equal to one is
equal to the total cluster density, $Q_1(v,t)=\pvt$.

Consider a cluster of velocity $v$ which contains at least $m$
particles.  Let us number the consecutive particles in a cluster from
right to left by the index $i$ and denote the rightmost particle as
$i=0$.  Denote the initial distance between the \ith\ and $(i-1)^{\rm
th}$ particle as $x_i$, as illustrated in Fig.~2.  We first solve for
$Q_{m=2}(v,t)$ and then generalize to any $m$.  Since $Q_2(v,t)$ is the
probability that a cluster of velocity $v$ has at least two particles at
time $t$, it is equal to the product of the probability that the
particle $i=0$ has survived up to time $t$, $\pvt$, and the probability
that the cluster $i=1$ (whose mass may be larger than unity) collides
with the particle $i=0$ prior to time $t$.

For this collision to occur, the collision partner from the left ($i=1$)
must have a velocity larger than $v$ and the interval $x_1<(v_1-v)t$
must be free of other clusters.  The probability for this composite
event is simply the product of each individual event. Since an interval
of length $x_1$ is empty with probability $\exp(-x_1)$, the collision
probability is
$$
Q_2(t)=\pvt\int_{v}^{\infty} dv_1 P_0(v_1)
\int\limits_{x_1<(v_1-v)t}dx_1\exp(-x_1).\eqno(22)
$$
The fact that the $v_1$-particle cannot be slowed down by any other
particle before colliding with the $v$-particle is crucial in obtaining
the solution.

To derive $\qnvt$ for general $m$, the joint velocity distance
distribution $P_0(v_i)\exp(-x_i)$ is integrated over the position and
velocity of the \ith\ particle for $i=1,\ldots,m-1$.  To ensure a
collision, all $m-1$ particles have to move faster than the lead
particle and the distance of the \ith\ particle from the lead particle
must obey $x_1+\cdots+x_i\le (v_i-v)t$.  Imposing these constraints on
the integration over the velocity and distance of the $n-1$ trailing
particles yields the formal exact expression for the cumulative
mass-velocity distribution,
$$
\qnvt=\pvt\prod_{i=1}^{m-1} \int_{v}^{\infty}dv_i P_0(v_i)
\int\limits_{x_1+\cdots+x_i<(v_i-v)t}dx_i \exp(-x_i).
\eqno(23)
$$
For the initial velocity distribution given by Eq.~(4),
we find the following asymptotic behavior (see Appendix A)
$$
\qnvt\simeq t^{\b-\a}a V^\m \exp\lr [ {-b (V+M)^{\m+2}} ],
\eqno(24)
$$
in terms of the scaling variables $M=m/t^{\a}$ and $V=vt^{\b}$.

In the long time limit, the joint mass-velocity distribution,
$\pnvt=\qnvt-Q_{m+1}(v,t)$ can be approximated by $P_m\approx -\partial
Q_m/\partial m$.  Performing the differentiation gives,
$$
\pnvt\simeq t^{\b-2\a}a\,b\, (\m+2)V^\m (V+M)^{\m+1}
\exp\lr [ {-b (V+M)^{\m+2}} ] ,
\eqno(25)
$$
which has been explicitly written in the asymptotic scaling form of
Eq.~(7).  This result provides a complete description of the traffic
aggregation process. It may be considered as the ballistic counterpart
of the well-known result [17] for diffusion-controlled aggregation in
one dimension.

For arbitrary $\mu$ we are unable to evaluate the integral over the
velocity and obtain the explicit mass distribution.  However, for the
particular case of a uniform initial velocity distribution, $\m=0$, it
is straightforward to show that
$$
\pn=a \exp(-a M^2/2)\qquad\m=0.
\eqno(26)
$$
Another amusing feature of the joint distribution function $\pnv$ for
$\m=0$ is the symmetry with respect to the variables $V$ and $M$.  Thus
the cluster mass distribution, $\pn$, and the cluster velocity
distribution, $\pv$, are identical Gaussian functions.

Generally, we are able to extract only asymptotic behavior from
Eq.~(23). However, in the special case of exponential distribution,
$\pvi=e^{-v}$, one can perform all the integrations and obtain an
explicit solution as detailed in Appendix B.

\bigskip\centerline{\bf V. Clustering in Heterogeneous Traffic Flow}

The above approach can be generalized to the case of a spatially
heterogeneous initial velocity distribution, $\pxvi$.  For simplicity,
we ignore the masses of clusters and limit ourselves to studying the
velocity distribution.  This time and space dependent velocity
distribution, $\pxvt$, may be found by a straightforward generalization
of the approach developed in section III for the spatially homogeneous
case.  The resulting expression for $\pxvt$ reads $$
\pxvt=P_0(x-vt,v)
\exp\lr [ {-\int_0^v dv'\int_{x-vt}^{x-v't} dx'P_0(x',v')} ] .
\eqno(27)
$$

As an illustrative example of the effects of a heterogeneous initial
particle distribution, consider the one-sided distribution in which
particles are placed with a fixed density to the left of the origin and
there are no particles to the right.  Thus $\pxvi=\theta(-x)\pvi$, with
$\theta$ the Heaviside step function.  For this initial distribution,
Eq.~(27) yields
$$
\pxvt=\theta(vt-x) \pvi
\exp\lr [ {-t\int_{x/t}^v dv'(v-v')P_0(v')-(vt-x)\int_0^{x/t} dv'P_0(v')} ] .
\eqno(28)
$$

In the long-time limit, the average velocity decays as $t^{-\b}$ and
hence the front propagates as $x=vt \sim t^\a$. Since the velocity and
the position of the front scale as power laws in time, Eq.~(28) can be
expected to have a scaling form.  Indeed, by introducing the scaled
variables $X=x/t^\a$ and $V=vt^\b$, one can recast Eq.~(28) into the
scaling form
$$
\pxvt \simeq t^{\b-\a}\Psi(X,V),
\eqno(29)
$$
with the scaling function
$$
\Psi(X,V)=\theta(V-X) aV^\m
\exp\lr [ {-b(V^{\m+2}-X^{\m+2})} ].
\eqno(30)
$$

When $X=0$, this scaling function coincides with Eq.~(16), the velocity
distribution for the homogeneous case, $\Psi(0,V)=\psi(V)$.  Notice also
that the density of clusters at scaled position $X$, $\cxt$,  equals
$$
\cxt=\ct \int_X^{\infty}dV\Psi(X,V)\bigg /\int_0^{\infty}dV\psi(V),
\eqno(31)
$$
with $\ct$ given by Eq.~(18). In the large $X$ limit, Eqs.~(30) and (31)
yield
$$
\cxt \simeq {\m+1 \over X} t^{-\a}, \quad {\rm for\ \ } X \gg 1.
\eqno(32)
$$

Consider now the total number of clusters that infiltrate the initially
empty positive half-line, $N(t)\equiv t^{\a}\int_0^{\infty}\cxt\,dX$.
The asymptotic behavior of this quantity is actually determined by the
finite upper cutoff of the integral, which in turn is given by the
position of the rightmost particle.  For such particles the velocity is
of order unity and hence $X_{\rm upper}=x_{\rm
max}/t^\a\sim vt/t^\a \sim t^\b$.  Therefore
$$ N(t) \simeq \int ^{t^\b}
{\m+1 \over X} dX \simeq \a \log(t).
\eqno(33)
$$
Thus the number of clusters entering the empty half-line grows only
logarithmically with time for arbitrary initial velocity distributions.
The only dependence on $\pvi$ in Eq.~(33) is the prefactor
$\a=(\m+1)/(m+2)$.

\bigskip\centerline{\bf VI. Clustering in Traffic Flow with Input}

In this section, we investigate traffic clustering when there is a
spatially uniform input of cars.  This generalization is motivated by
 real traffic where cars may
enter and exit a roadway.  For the specific case of a spatially
homogeneous input of cars we can determine the velocity distribution
using techniques similar to those employed for the traffic coalescence
model with no input.

Denote by $R(v,t)$ the input rate of particles with velocity $v$ at time
$t$ per unit length.  The velocity distribution function for this
system, $\pvt$, can be expressed as a convolution of the flux and the
the probability that a particle which was injected at time $t'$
maintains its velocity $v$ up to time $t$ in the presence of the input,
$S_I(v,t,t')$,
$$
\pvt=\int_0^t dt' \, R(v,t')\, S_I(v,t,t'). \eqno(34)
$$
In writing this expression, we have assumed that the system is initially
empty.  A particle which was injected at time $t'$ will survive until
time $t$ if it avoids collisions with all slower particles which were
present in the system at time $t'$, as well as avoids collisions with
all particles which are injected at later times $t''>t'$.  The
probability for this composite event is simply the product of the
probabilities of each event,
$$
S_I(v,t,t')=\exp\Biggl[-\int_0^v dv'(v-v')
\left\{P(v',t')(t-t')+\int_{t'}^t dt'' R(v',t'')(t-t'')\right\}\Biggr] .
\eqno(35)
$$
Here the first factor, obtained from Eq.~(13) by replacing $P_0(v')$
with $P(v',t')$, yields the probability of avoiding collisions with
particles injected prior to time $t'$.  The second factor represents the
product of exclusions of the type off Eq.~(12) for times larger than
$t'$ and velocities smaller than $v'$.  This factor accounts for the
probability that there are no collisions with particles injected at
times $t'', t''>t'$. Note that the kernel of the second factor involves
the input rate $R$ at time $t''$, which plays the role of the initial
velocity distribution at this instant of time.  Substituting Eq.~(34)
into Eq.~(35) gives a nonlinear integral equation that describes the
kinetics of traffic clustering in the presence of homogeneous particle
input.  In the following, we will assume that the flux is constant {\it
both} in time and space, $\rvt=\pvi$ with $\int_0^{\infty} dv\pvi=1$, so
that the governing integral equation is $$
\pvt=\pvi\int_0^t dt'
\exp\lr [ {- {1 \over 2}t'^2\int_0^v dv'(v-v')\pvi
-t'\int_0^v dv'(v-v')P(v',t-t')} ] .
\eqno(36)
$$

In parallel with the case of no input, consider again an initial
velocity distribution with a power-law small-velocity tail, $\pvim$.
Furthermore, let us assume that the concentration and velocity continue
to vary as $\cta$ and $\vtb$, respectively, and that the velocity
distribution continues to have the scaling form of Eq.~(9), $\pvt \sim
t^{\b-\a}\pv$. Substituting these into Eq.~(36), one can extract
consistency conditions for the exponents $\a$ and $\b$. For example, the
powers of time on both sides of Eq.~(36) should be equal -- hence
$\b-\a=-\b\m+1$.  Furthermore, both terms in the exponential in the
right-hand side of Eq.~(36) cannot depend on time explicitly -- hence
$\b(\m+2)=2$ and $\a+\b=1$, respectively. These conditions yield
$\a=\m/\mt$ and $\b=2/\mt$.  Notice, however, that when $\m$ is positive
the exponent $\a$ is also positive and therefore the concentration,
$\cta$, decays to zero.  This is in obvious contradiction with the
nature of the problem: with a constant flux the concentration may grow
indefinitely or a steady state concentration may be reached.  Thus, one
can expect that the above description holds only for $\m<0$.  For
positive $\m$, we anticipate that the system reaches a steady state with
a constant concentration and a typical cluster mass $n\sim t$. At the
transition $\m=0$, a logarithmic temporal dependence is anticipated to
occur.

While we cannot confirm the above picture rigorously, we can provide
heuristic justification.  First {\it assume} that the velocity
distribution evolves towards a steady state $\pinf$. Then as $t \to
\infty$ Eq.~(36) becomes
$$
\pinf=\pvi\int_0^{\infty} dt'
\exp\lr [ {-{1 \over 2}t'^2\int_0^v dv'(v-v')\pvi
-t'\int_0^v dv'(v-v')P_{\infty}(v')} ] .
\eqno(37)
$$
If $P_0(v)\sim v^\m$ as $v\to 0$,  Eq.~(37) suggests a similar
behavior for the steady state velocity distribution function,
$$
\pinf \simeq Av^{\nu} \quad v \ll 1.
\eqno(38)
$$
Since the concentration of clusters tends to the steady state limit
$\cinf=\int_0^{\infty}\pinf\,dv$, the exponent $\nu$ must satisfy the
inequality $\nu>-1$.  If one substitutes the assumed power law behaviors
for $\pvi$ and $\pinf$ into Eq.~(37), three possibility arise in the
limit $v \to 0$ which depend on the sign of $\nu-{\m \over 2}+1$.  In the
case where $\nu>{\m \over 2}-1$, the first exponential factor in
Eq.~(37) provides the dominant contribution.  However, a simple
calculation of the integral shows that $\nu={\m \over 2}-1$.  Similarly,
for $\nu<{\m\over 2}-1$, one again finds $\nu={\m \over 2}-1$.  Only the
last possibility, $\nu={\m \over 2}-1$, appears to be self-consistent.
Since $\nu>-1$, we obtain $\m>0$.  Therefore starting from the
assumption that the system reaches the steady state we have obtained
that the exponent $\m$ should be positive.  This provides evidence for
our conclusion that $\m=0$ demarcates the scaling and steady state
behaviors.

Notice also that for $0<\m \ll 1$, an asymptotic analysis of Eq.~(37)
gives the numerical prefactor in Eq.~(38), $A \simeq \sqrt{a\m/2}$.
This yields the estimate for the steady state concentration of clusters,
$$
\cinf \simeq \sqrt{2a \over \m} \quad  0<\m \ll 1,
\eqno(39)
$$
Since $\cinf$ diverges as $\m \to 0$ this indicates that at the critical
value $\m, \m=0$, the system is still evolving.

Assuming the scaling form, obtained by the power counting analysis
of Eq.~(36), let us examine the asymptotic
behavior of the velocity distribution and the typical concentration.  If
we substitute the scaling assumptions
$$
\pvt \simeq t^{\b-\a}\pv \quad{\rm with}\quad \a={\m \over \m+2},
\quad \b={2 \over \m+2},
\eqno(40)
$$
into Eq.~(36), we arrive at the following equation for the scaling
function $\pv$,
$$
\pv=aV^\m \int_0^1d\tau
\exp\lr[ {-{1 \over 2}b\tau^2 V^{\m+2} -
\tau(1-\tau)^{\b-\a}\int_0^V\psi[V'(1-\tau)^\b](V-V')dV'} ],
\eqno(41)
$$
where $\tau=t'/t$ and $b=a/(\m+1)(\m+2)$.

Although we are unable to solve this nonlinear integral equation in
general, we can obtain interesting information regarding the interesting
borderline case of $\mu=0$.  This case corresponds to $\a=0$, suggesting
that the concentration grows slower than algebraically in time.
However, the concentration of clusters is given by
$$
\ct \simeq t^{-\a}\int\pv dV,
\eqno(42)
$$
and for $\mu=0$ the integral diverges at the upper limit.  To obtain the
asymptotic behavior we consider Eq.~(41) for the case
$\mu=0$,
$$
\pv=a\int_0^1d\tau
\exp\lr[ {-{1 \over 2}b\tau^2 V^2 -
\tau(1-\tau)\int_0^V\psi[V'(1-\tau)](V-V')dV'} ].
\eqno(43)
$$ If we temporarily ignore the second term in the exponent, we find
$\pv \sim V^{-1}$ for $V \to {\infty}$.  If we then include the second
term in the exponent and apply the previous asymptotic behavior of
$\pv \sim V^{-1}$, we find $\pv \sim
(V\log{V})^{-1}$. These estimates suggest the ansatz
$$
\pv \simeq C V^{-1} (\log V)^{-\l},
\eqno(44)
$$
for $V \gg 1$.  Upon substituting Eq.~(44) into Eq.~(43) one obtains the
constants $\l=1/2$ and $C=1/\sqrt{2}$.  With these values, the integral
in the right-hand side of Eq.~(42) diverges as $\sqrt{2\log V}$, where
$V$ now denotes the maximum value of the scaled velocity.  Since $\beta=1$,
this maximal velocity is proportional to $t$, which therefore
suggests the logarithmic time dependence
$$
\ct \simeq \sqrt{2\log t}.
\eqno(45)
$$
In the complementary case of $\m<0$, we have confirmed that the naive
scaling ansatz is consistent with Eq.~(42).

\bigskip\centerline{\bf VII. Conclusion}

We have introduced a simple ballistic aggregation process that mimics
the kinetics of clustering in a single lane of traffic.  Through
direct probabilistic approaches, the analytical forms of the cluster
velocity distribution and the joint mass-velocity distribution have been
derived.  For an initial velocity distribution of the form $P_0(v)\sim
v^\mu$ as $v\to 0$, both the average velocity and the average cluster
mass have power law time dependences with exponents that are rational
functions of $\mu$.  This qualitative behavior is similar to that
observed in the closely related ballistic annihilation process.  We are
also able to determine the asymptotic form of the joint mass-velocity
distribution.

Our model can also be analyzed in the cases of a spatially
heterogeneous particle distribution and continuous input of particles.
For the simple case of an initial one-sided spatial distribution, the
system evolves towards a scaling distribution both in velocity and
spatial variables.  We have thus found that the total number of clusters
in the initially empty half-line grows logarithmically with time for all
initial velocity distributions.  When there is a steady input of
particle in an initially empty system, we have found that there is a
transition between steady state behavior for $\m>0$ and transient
behavior for $\m<0$ which is similar to that found when there is no in
put.  For the borderline case of the uniform distribution, $\m=0$, we
have found that the total concentration of clusters grows as $\sqrt{\log
t}$.

The irreversible traffic model introduced in this paper leads to
ever-growing clusters.  To describe traffic flows more realistically,
several mechanisms to induce a steady state can be envisioned.  For
example, the input model can be generalized to incorporate a flux out
of the system.  Another realistic direction is to allow a faster car to
pass a slower car at a rate which is some increasing function of the
velocity difference of the two cars. This would allow a fast car to
traverse a cluster car-by-car and ultimately regain its intrinsic
velocity once the cluster is completely passed.  It may prove
interesting to examine the steady-state properties for this class of
models.

\bigskip\beginsection\centerline{Acknowledgements}

One of us (SR) thanks Chris Myers for useful discussions.  We also
gratefully acknowledge ARO grant \#DAAH04-93-G-0021, NSF grant
\#DMR-9219845, and the Donors of The Petroleum Research
Fund, administered by the American Chemical Society, for partial support
of this research.

\vfill\eject
\centerline{\bf Appendix A: Derivation of Eq.~(24)}

In this appendix, we derive the asymptotic form of the cumulative
velocity-mass distribution which is valid for an arbitrary initial
velocity distribution with a power-law small velocity tail.  The
starting point for calculation of $\qnvt$ is the formal expression of
Eq.~(23). By interchanging the order of the velocity and spatial
integrations, the expression can be rewritten as
$$
\qnvt=\pvt\prod_{i=1}^{m-1}
\int_0^\infty dx_i \exp(-x_i)
\int_{v+(x_1+\cdots+x_i)/t}^{\infty}dv_i\, P_0(v_i).
\eqno({\rm A}1)
$$
Since the details of the initial distribution for large velocities do
not change the form of the scaling solution, we may treat the more
general power-law case by choosing a specific initial distribution of
velocities whose form is convenient for performing the integration over
the velocity in the right-hand side of Eq.~(A1).  Since the velocity
distribution near $v=0$ has a power-law tail, the most convenient
initial distribution is
$$
P_0(v)=a v^\m \exp \lr ( {-a v^{\m+1}/\mo} ) .
\eqno({\rm A}2)
$$
For this initial distribution, the integration over the velocity
variables is immediate and one finds,
$$
\qnvt=\pvt\prod_{i=1}^{m-1}
\int_0^\infty dx_i \exp(-x_i)
\exp\lr [ {-a\Bigl( v+(x_1+\cdots+x_i)/t \Bigr)^{\m+1}/\mo } ] .
\eqno({\rm A}3)
$$

To evaluate the multiple spatial integral, we define
$f(z)\equiv\exp\left(-a z^{\m+1}/\mo\right)$ and expand $f(z)$ to first
order about the point $z=v$ and then exploit a number of simplifications
associated with performing the integrals over the factors $e^{-x_i}$.
This gives
$$
\eqalign{
Q_m(v,t)&=\prod_{i=1}^{m-1} \int_0^\infty dx_i\, e^{-x_i}
f\bigl(v+\e(x_1+\cdots+x_i)\bigr)\cr
&\approx\prod_{i=1}^{m-1} \int_0^\infty dx_i\, e^{-x_i}
f(v)+\bigl(\e(x_1+\cdots+x_i)\bigr)f'(v),\cr
&\approx\prod_{i=1}^{m-1} f(v+\e i) +O(\e^2),\cr}
\eqno({\rm A}4)
$$
where $\e=1/t$.  By substituting the explicit functional form,
$f(z)=\exp\bigl(-az^{\m+1}/\mo\bigr)$, we have
$$
\qnvt=\pvt\prod_{i=1}^{m-1} \exp\lr( {-a(v+i/t)^{\m+1}/\mo} ).
\eqno({\rm A}5)
$$
Finally, the product of the exponential factors is written as an
exponent of a sum. In the limit of large $m$, this sum is equivalent to
the integral $a\int_0^{m-1} dy (v+y/t)^{\m+1}/\mo$.  Evaluating this
integral, the asymptotic form of the cumulant mass-velocity
density is obtained as
$$
\qnvt \simeq a v^\m\exp\lr [ {-b(vt^\b+mt^{-\a})^{\m+2}} ] .
\eqno({\rm A}6)
$$
In evaluating this asymptotic expression, the exponential factor of
$\pvt$ cancels the factor that emerges from lower limit of the
integration over $y$.
In Eq.~(24), the above expression is written as a scaling function.

\bigskip\centerline{\bf Appendix B: Analytical Solution for the Exponential}
\centerline{\bf Initial Velocity Distribution}

We outline here the explicit analytical solution for $Q_m(v,t)$ for the
exponential initial velocity distribution, $\pvi=e^{-v}$, which
corresponds to the special case $\m=0$ and $a=1$.  For this exponential
distribution, the integration of $P_0(v_i)$ over the velocity variables
$v_i$, according to Eq.~(A1), equals
$\exp[-\bigl(v+(x_1+\cdots+x_i)/t\bigr)]$.  Hence, we obtain
$$
\qnvt=\pvt e^{-(m-1)v}\prod_{i=1}^{m-1}
\int_0^\infty dx_i \exp\biggl[-x_i\biggl(1+{m-i\over t}\biggr)\biggr].
\eqno({\rm B}1)
$$
Upon integration over the space variables, the following exact
expression is found for $\qnvt$,
$$
\qnvt=\pvt e^{-(m-1)v} t^{m-1} {\Gamma(t+1)\over \Gamma(t+m)},
\eqno({\rm B}2)
$$
where $\Gamma(z)=\int_0^{\infty} x^{z-1}\exp(-x)$ is the Euler
gamma function and the velocity distribution obtained from
Eq.~(14) is $\pvt=\exp\bigl[-v-t\bigl(e^{-v}-1+v\bigr)\bigr]$.

The exact forms for the joint mass-velocity distribution
and for the mass distribution can be evaluated from Eq.~(B2)
by taking the appropriate limits $m \to \infty$ and $v \to 0$.
The resulting asymptotic expressions are identical with
the expressions of Eqs.~(25) and (26) respectively.

\vfill\eject
\bigskip\beginsection\centerline{References}

\refi See, \eg, {\sl Transportation and Traffic Theory},
edited by N.~H.~Gartner and N.~H.~M.~ Wilson (Elsevier, New York, 1987);
W.~Leutzbach, {\sl Introduction to the Theory of Traffic Flow}
(Springer-Verlag, Berlin, 1988).
\refi O.~Biham, A.~Middleton, and D.~Levine, \pra 46 6124 1992 .
\refi T.~Nagatani, \pre 48 3290 1993 .
\refi K.~Nagel and M.~Schreckenberg, \jdep 2 2221 1992 .
\refi A.~Schadschneider and M.~Schreckenberg, \jpa 26 L679 1993 .
\refi T.~Nagatani, \jpa 26 L781 1993 .
\refi I.~Prigogine and R.~Herman, {\sl Kinetic Theory of Vehicular Traffic}
(Elsevier, New York, 1971).
\refi G.~B.~Whithem, {\sl Linear and Nonlinear Waves}
(Wiley, New York, 1974).
\refi B.~S.~Kerner and R.~Konhauser, \pre 48 2335 1993 .
\refi G.~F.~Carnevale, Y.~Pomeau, and W.~R.~Young, \prl 64 2913 1990 .
\refi Y.~Jiang and F.~Leyvraz, \jpa 26 L179 1993 .
\refi Y.~Elskens and D.~L.~Frisch, \pra 31 3812 1985 ; J. Krug and H.
Spohn, \pra 38 4271 1988 .
\refi E.~Ben-Naim, S.~Redner, and F.~Leyvraz, \prl 70 1890 1993 .
\refi R.~M.~Bradley and P.~N.~Strenski, \prb 40 8967 1989 .
\refi B.~Derrida, C.~Godr\`eche, and I.~Yekutieli, \pra 44 6241 1991 .
\refi Yu.~A.~Andrienko, N.~V.~Brilliantov, and P.~L.~Krapivsky,
\pra 45 2263 1992 ; P.~L.~Krapivsky, \jcp 97 8817 1992 .
\refi J.~L.~Spouge, \prl 60 871 1988 .

\bigskip
\centerline{\bf Figure Captions}
\medskip
{\parindent = 0.8 true in
\item{Figure 1.} Schematic illustration of the irreversible traffic
model.  A faster car overtakes a slower car and after the encounter, the
faster car assumes the velocity of the slower car.
{\parindent = 0.8 true in
\item{Figure 2.} Illustration of the initial configuration of a possible
three car cluster.

\vfill\eject\bye

\end